# Transport through modes in random media

Jing Wang[1] & Azriel Z. Genack[1]

[1]*Department of Physics, Queens College of the City University of New York,*

*Flushing, New York 11367, USA*

**Excitations in complex media are superpositions of eigenstates. These are variously referred to as levels for quantum systems and modes for classical waves. Though the Hamiltonian of complex systems may not be known or solvable, Wigner conjectured[1] that the statistics of energy level spacings would be the same as for the eigenvalues of large random matrices. This has explained key characteristics of neutron scattering spectra[2]. Subsequently, Thouless and coworkers argued[3,4] that the metal-insulator transition[4-6] could be described by a single parameter, the ratio of the average width and spacing of electronic energy levels. When this dimensionless ratio falls below unity, conductivity is suppressed by Anderson localization[5] of the electronic wave function. However, because of spectral congestion due to the overlap of modes[7-9], even for localized waves, the prospect of forging a comprehensive modal approach to wave propagation has not been realized. Here we show that the field speckle pattern[10] of transmitted microwave radiation can be decomposed into a sum of patterns of the modes of the medium. We find strong correlation between modal field speckle patterns which leads to destructive interference between modes. This allows us to explain complexities of steady state and pulsed transmission of localized waves and to harmonize wave and particle descriptions of diffusion.**

Modes of the field in media for which the corresponding particles freely diffuse extend throughout the sample. Energy is then readily transported to the margins of the sample where it leaks through the boundary. Since the mode lifetime is then short and its linewidth correspondingly broad, the average spectral width of modes exceeds the average spacing between neighbouring modes, $\delta\omega > \Delta\omega$[3,4,6]. Level repulsion of spectrally overlapping levels leads to universal fluctuations in conductance in small metallic samples[11] and is associated with classically chaotic motion[12] for waves inside a cavity. In contrast, the coupling of exponentially peaked modes to the environment surrounding the sample is exponentially small so that, $\delta\omega < \Delta\omega$[3,4,6,11]. Thus average transmission is then suppressed[3-7] while relative fluctuations of transmission[13] or conductance[14] are enhanced as a result of occasional resonant tunnelling through localized modes. The degree of spectral mode overlap, $\delta \equiv \delta\omega/\Delta\omega$, often referred to as the Thouless number, is thus a fundamental localization parameter. Anderson localization has been observed for electrons[15] and cold atoms[16], as well as for classical waves[14,17,18] such as light[19], microwave radiation[13], plasmons[20] and ultrasound[21,22].

A decisive step forward in the study of transport was taken with the development of the scaling theory of localization[4,6] in which the scaling of a directly measurable quantity, the dimensional conductance, $g$, was shown to depend only upon $g$ itself and the dimensionality of the system at zero temperature. It was argued that the dimensionless conductance, $g = G/(e^2/h)$, where $G$ is the electronic conductance, $e$ is the electron charge and $h$ is Planck's constant, is equal to the Thouless number, $\delta = g$ [4,6]. $g$ can also be defined for classical waves via the Landauer relation which equates the dimensionless conductance to the ensemble average of the transmittance, $<T>$. The transmittance is the sum of coefficients of total transmission for the $N$ propagating transverse incident channels, $a$, $g = \langle T \rangle = \left\langle \sum_{a=1}^{N} T_a \right\rangle$.

The statistics of transmission in nondissipative samples are also determined by $g$[13,23-26]. The variance of the total transmission normalized by its ensemble average, $\text{var}(s_a \equiv T_a/\langle T_a \rangle)$, and the variance of normalized intensity, $\text{var}(s_{ab}=T_{ab}/<T_{ab}>)$ are linked via the relation, $\text{var}(s_a) = [\text{var}(s_{ab})-1]/2 = 2/3g$ [25,26]. More generally, the full distribution of intensity and total transmission can be expressed in terms of a statistical conductance $g' \equiv 2/3\text{var}(s_a)$ [13,27], even in strongly scattering dissipative samples. In the absence of inelastic processes $g'$ reduces to $g$ [13].

Thus we expect that key aspects of propagation in nondissipative systems are linked to the statistical properties of the underlying modes via the relations, $\delta \sim g = g'$. Confirming this string of relations presents a challenge because these only hold in the absence of dissipation. Therefore measurements must either be carried out in nondissipative systems or a means must be found for obtaining these parameters from measurements in the presence of inelastic scattering, dephasing or absorption. Here we determine $\delta$ directly from the linewidths and spacings of modes. Since the linewidth of the $n^{th}$ mode, $\Gamma_n$, is the sum of the rate at which the modal energy leaks through the sample boundaries, $\Gamma_n^0$, and the absorption rate, $1/\tau_a$, the linewidth in an equivalent sample without absorption is given by, $\Gamma_n^0 = \Gamma_n - 1/\tau_a$. This makes it possible to compute $\delta$ and $g'$ in samples at two lengths in which the impact of absorption is eliminated.

Measurements of the frequency variation of the speckle pattern of microwave field transmitted through random samples of alumina spheres at low filling fraction contained in a copper tube are carried out as shown schematically in Fig. 1a. The field is detected with use of a short wire antenna connected to a vector network analyzer. Spectra are taken over a tight grid of points on the sample output. Intensity speckle patterns at a fixed frequency such as shown in the figure are formed from these spectra. New sample configurations are realized by momentarily rotating the sample tube about its axis. Over an ensemble of 200 sample configurations for samples of length 61 cm in the frequency range 10-10.24 GHz, we find, $g' = 0.23$. This is well above the value at the localization threshold of unity in nonabsorbing samples and corresponds to a sample of length of more than twice the localization length in our sample with absorption rate $1/\tau_a = 0.0064$ ns$^{-1}$.

Since energy leaks from the open ends of the sample tube and radiation is absorbed by the sample, electromagnetic energy is not conserved and the time-evolution operator is not hermitian. Nonetheless, we expect that the time evolution of the wave following pulsed excitation can be expressed as a sum of terms, each of which is a product of a volume field speckle pattern and an exponentially decaying sinusoidal function. These terms, which are often referred to as quasi-normal modes, can form a complete set for outgoing waves[28]. In the frequency domain, the field at the output surface can be expressed as a superposition of such modes,

$$E_j(\mathbf{r},\omega) = \sum_n a_{n,j}(\mathbf{r}) \frac{\Gamma_n/2}{\Gamma_n/2 + i(\omega - \omega_n)} = \sum_n a_{n,j}(\mathbf{r}) \varphi_n(\omega). \tag{1}$$

Here, $a_{n,j}(\mathbf{r})$ is the spatial variation of the $j^{th}$ polarization component of the field for the $n^{th}$ mode with central frequency $\omega_n$ and linewidth $\Gamma_n$ given in radians. $\varphi_n(\omega)$ is the frequency variation of the mode given by the Fourier transform of $\exp(-\frac{\Gamma_n}{2})\cos\omega_n t$ for $t > 0$. The average number of modes per configuration over the frequency range studied is 47. Since spectra at all $r$ in a given configuration share a common set of $\omega_n$ and $\Gamma_n$, the mode expansion in equation (1) can be fit simultaneously to spectra at many points to give $\omega_n$ and $\Gamma_n$ and the corresponding mode speckle patterns. Measured intensity spectra at two points and spectra obtained from the simultaneous modal fit of spectra at 45 points are shown in Fig. 1b. The measured normalized total transmission, $s_a$, and the sum over the output surface of $|E_j(\mathbf{r},\omega)|^2$ obtained from this global fit are shown in Fig. 2a. Excellent agreement between measurements and the global fit is obtained for both local (Fig. 1b) and integrated transmission (Fig. 2a).

To explore the contribution of modes to total transmission, we consider a narrow frequency range around the strong peak at 10.15 GHz for the same configuration for which intensity spectra are shown in Fig. 1b. The asymmetrical shape for the line in both intensity and total transmission indicates that more than a single mode contributes to the peak. The modal analysis of the field spectra shows that three modes contribute substantially. Total transmission spectra for each of the three modes closest to 10.15 GHz are plotted in Fig. 2a. The integrated transmission for two of these modes corresponding to the 28th and 29th mode in the spectrum starting at 10 GHz, are each greater than for the measured transmission peak indicating that these modes interfere destructively. The intensity and phase patterns for each of these modes are shown in Figs. 2b-e. Aside from a difference in average transmission, the intensity speckle patterns of the two modes are nearly identical. At the same time, the distributions of phase shift at 10.15 GHz for the two modes are similar up to a nearly constant phase difference with average value $\Delta\varphi = 1.02\ \pi$. The surprising similarity between the speckle patterns for these overlapping modes suggests that these spectrally overlapping modes are formed from coupled resonances which overlap spatially within the sample[7-9]. The similarity in the intensity speckle patterns of adjacent modes and the uniformity of the phase shift across the patterns of these modes makes possible interference between modes across the entire speckle pattern.

The modal decomposition of field spectra affords a full account of dynamic as well as of steady-state transmission. Pulsed transmission is obtained by taking the Fourier transform of the product of transmitted field spectra $E_j(\mathbf{r},\nu)$ and the Fourier transform of an incident Gaussian field pulse, $E_j(r,\nu)\exp(-(\nu-\nu_0)^2/2\sigma^2)$. From this, we compute the time-frequency spectrogram of the total transmission, $T_a(t, \nu_0)$, corresponding to the variation with carrier frequency $\nu_0$ of the sum of intensity over all points on the output surface on which measurements of field are made at delay time $t$ from the peak of an incident Gausian pulse, $T_a(t, \nu_0) = \Sigma_\mathbf{r}|E(\mathbf{r}, t, \nu_0)|^2$. $T_a(t, \nu_0)$ is indicated by the colour scale in the x-y plane in Fig. 3. The evolution of the spectra of total transmission is further indicated by plotting $T_a(t, \nu_0)$ normalized to the average over each spectrum for four delay times. The increasing impact of long-lived narrow-line modes is manifest in the decreasing number of surviving modes with substantial relative intensity. This results in an increasing variance of normalized transmission with time delay. The variation with time of the spectrogram of total transmission normalized by the peak transmission for the mode with center frequency $\nu_n$ = 10.04352 GHz is seen in Fig. 1b to narrow and take on the Gaussian lineshape of the incident pulse. Figure 3c shows the decay of this peak. The decay rate of 1.18 $\mu s^{-1}$ is essentially equal to the linewidth $\Gamma_n$ = 1.29 MHz.

The average temporal variation of total transmission, $\langle T_a(t) \rangle$, is obtained by integrating the time-frequency spectrogram over frequency in each configuration at different times and averaging over all configurations. The progressive suppression of transmission in time by absorption may be removed by multiplying $\langle T_a(t) \rangle$ by $\exp(t/\tau_a)$ to give, $\langle T_a^0(t) \rangle = \langle T_a(t) \rangle \exp(t/\tau_a)$, in which the loss of energy is due only to leakage from the sample. The same result is obtained by transforming into the time domain using spectra computed from the modal decomposition of the field in which the impact of absorption on each mode is eliminated. The decay of $\langle T_a^0(t) \rangle$, shown as the solid curve in Fig. 4, is seen to slow considerably with time delay[22,29]. This reflects a broad range of modal decay rates from long-lived localized modes peaked at points remote from the sample surface[3,4,14] to more extended short-lived modes that overlap in space and frequency[7-9].

The measured decay is compared to the incoherent sum of transmission for all modes in the random ensemble, $\langle \sum_n T_{an}^0(t) \rangle$, shown as the dashed curve in Fig. 4. $\langle \sum_n T_{an}^0(t) \rangle$ is substantially larger than $\langle T_a^0(t) \rangle$ at early times but then converges to $\langle T_a^0(t) \rangle$. Though transmission associated with individual modes rises with the incident pulse, transmission at early times is strongly suppressed by destructive interference of modes with strongly correlated field speckle patterns such as those shown in Fig. 2. The delayed rise of pulsed transmission seen in Fig. 4 is a remnant of particle diffusion associated with transport in weakly scattering samples and reflects the low probability of particles traversing the sample by a sequence of scattering events all in the forward direction. At later times, the random frequency differences between modes leads to additional random phasing between modes, $(\omega_{n+1} - \omega_n)t$. This leads to a peak in transmission after a delay time comparable to the inverse of the typical mode linewidth. By this time, destructive interference is substantially diminished and subsequently averaged pulsed transmission approaches the incoherent sum of decaying modes.

In order to determine $\delta$ and relate it to other localization parameters which encapsulate different aspects of propagation, we need to consider propagation in samples without absorption. We find in one-dimensional simulations that the phases of $a_{n,j}(\mathbf{r})$ along the length of the sample are barely affected by absorption and that spectra found without absorption are replicated to high accuracy by substituting $a_{n,j}(\mathbf{r})\Gamma_n/\Gamma_n^0$ for $a_{n,j}(\mathbf{r})$ and $\Gamma_n^0 = \Gamma_n - 1/\tau_a$ for $\Gamma_n$ in equation (1). Making this substitution in the parameters found from the modal fit to measured spectra, we find spectra that would be obtained in the absence of absorption.

Once the impact of absorption is removed, $\delta$ may be defined purely in terms of the statistical properties of modes. We identify the average level width, or inverse Thouless time, with the average modal leakage rate, $\delta\omega \equiv \tau_{Th}^{-1} \equiv \langle 1/\Gamma_n^0 \rangle^{-1}$, and equate the average level spacing $\Delta\omega$ to the average angular frequency difference between neighbouring modes in 200 sample configurations to give, $\delta = \delta\omega/\Delta\omega = 0.17$. This equals the statistical conductance, $g' = 0.17$, determined from the variance of intensity in spectra in which absorption is removed. $\delta$ and $g'$ are also found to be close for an ensemble of 40 samples realizations of length 40 cm with, $\delta = 0.43$ and $g' = 0.39$. Figure 5 presents the relationship between two measures of localization; the inverses of $\delta$ and $g'$. The origin of the plot is the diffusive limit in which the number of modes encompassed within the linewidth diverges and fluctuations of total transmission and the degree of fractional intensity correlation vanish[23-26]. The red line with slope unity is an extension of the equality of $\delta$ and $g'$ from the diffusive[25,26] to the localized[13,27] regime.

We have shown that steady state spectra and wave dynamics are explained in terms of destructive interference which is a consequence of strong field correlation across transmitted modal speckle patterns. Notwithstanding the complex correlation found between modes, the statistics of transmission are simply related to the mode overlap parameter, $\delta$, as seen in the equality of $g'$ and $\delta$ for diffusive and localized waves. Mode statistics are crucial for nonlinear as well as for linear phenomena and determine the threshold for narrow-line lasing in random amplifying media in systems in which linewidths and level spacings are comparable[30]. The results presented here show that the road to a full understanding of transport in random systems passes through standing waves of the medium.

**Methods summary**

Measurements are carried out of the microwave field transmitted through an ensemble of samples of randomly positioned alumina spheres embedded in Styrofoam spheres contained within a copper tube. The tube is 99.999% copper to minimize losses in reflection. It has a diameter of 7.2 cm and supports 30 transverse modes of the empty waveguide over the frequency range of the experiment of 10-10.24 GHz. The sample is composed of 99.95% alumina spheres of diameter 0.95 cm and refractive index 3.14 embedded in Styrofoam spheres of diameter 1.9 cm and refractive index 1.04 to achieve an alumina volume filling fraction of 0.068. Waves are thoroughly mixed in the quasi-one-dimensional sample geometry of the experiments so that intensity statistics are essentially uniform across the output. The signal is detected by a 4-mm long 300-μm diameter wire antenna translated on a 2-mm square grid over the output surface. The separable nonlinear least squares fitting method was used to simultaneously decompose field spectra at each of 45 positions into a sum over modes using a common set of $\omega_n$, $\Gamma_n$. The minimum is found for the sum, $\min \chi^2 = \min \sum_{p=1}^{P=45} \left| E(r_p, \nu) - \sum_{n=1}^{N} a_n(r_p) \varphi_n \right|^2$, where $r_p$ are the different positions whose spectra are fit. The problem is solved to yield the $\omega_n$, $\Gamma_n$ and the mode speckle patterns by using classical nonlinear least squares optimization methods, such as the Levenberg-Marquardt algorithm. Figures 3 and 4 are computed from the measured field spectra using time-frequency analysis for pulses with $\sigma = 0.85$ MHz and 3 MHz, respectively, corresponding to Gaussian temporal pulse widths of $\sigma_t = 187.4$ ns and 53.1 ns.

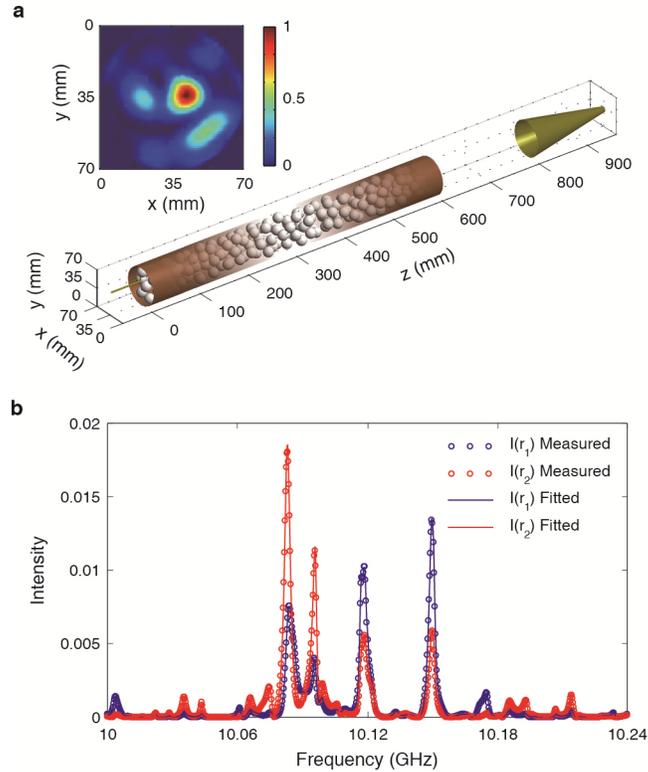

**Figure 1 | Measurements of transmission through random media. a,** Microwave radiation is launched from a horn placed 20 cm before the sample. Field spectra of transmission through randomly positioned alumina spheres contained in a copper tube are measured at points on a 2-mm grid over the output surface using a short wire antenna and a vector network analyzer. Squaring the field at each point gives the intensity speckle pattern. **b,** Intensity spectra at two detector positions and the fit of equation (1) to the data.

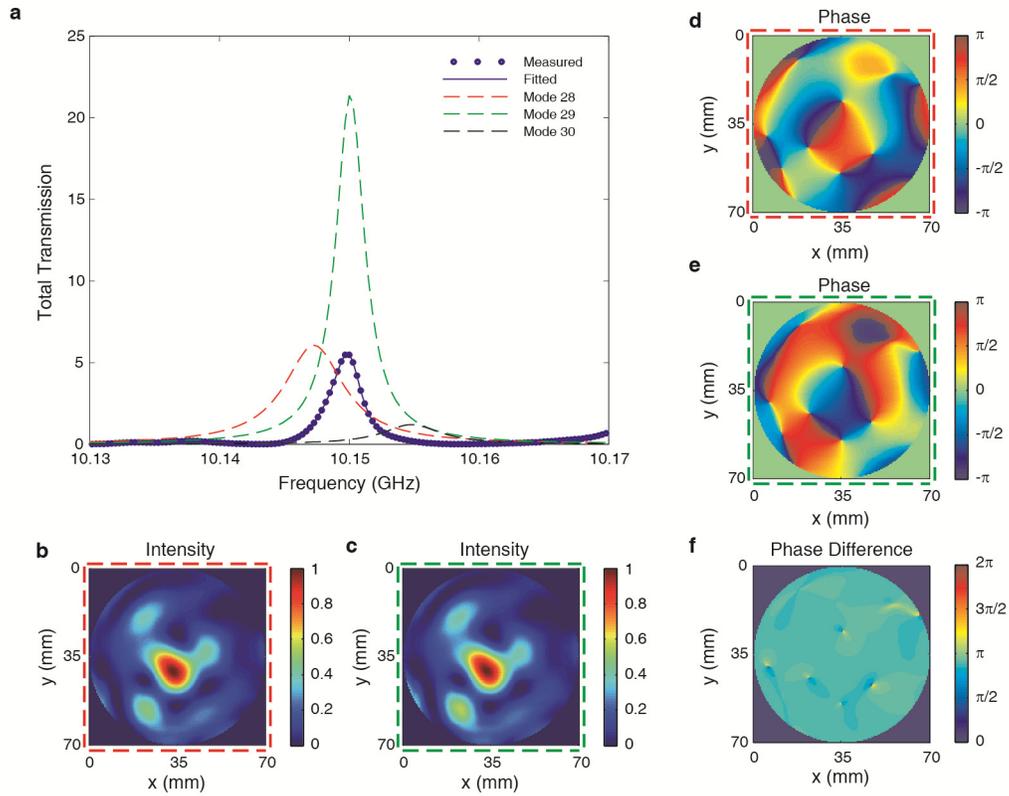

**Figure 2 | Total transmission and mode speckle patterns. a,** Three modes contribute to the asymmetric peak in the total transmission spectrum (blue) for the same configuration as in Fig. 1, #28 (red), #29 (green) and #30 (black). **b, c,** Dashed red and green lines surrounding intensity speckle patterns for modes #28 and #29, respectively. **d, e,** Phase patterns for these two modes. **f,** The phase in Mode #29 is shifted by nearly a constant of $1.02\pi$ rad relative to Mode #28.

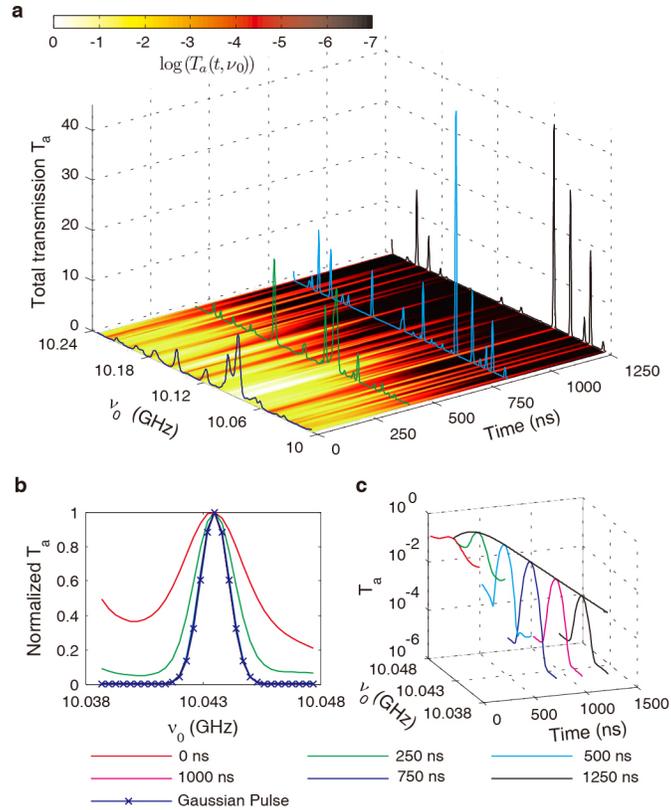

**Figure 3 | Time-frequency analysis. a,** Logarithm of time-frequency spectrogram of total transmission plotted in colour-scale in the *x-y* plane. The central frequency, $\nu_0$, of the incident Gaussian pulse of linewidth, $\sigma$ = 0.85 MHz, is scanned. Each of the four spectra of total transmission at different delay times are normalized to its average. **b,** Spectrograms of total transmission for the mode with center frequency, $\nu_n$ = 10.04352 GHz are normalized by the peak transmission for different delay times. The colour of the curves in b and c indicate different delay times for the spectrograms in b and c as designated in the legend at the bottom of the figure. All the curves overlap for delays of 500 ns and longer. **c,** The decay rate of the peak in part b of 1.18 µs$^{-1}$ is essentially equal to the linewidth $\Gamma_n$ = 1.29 MHz.

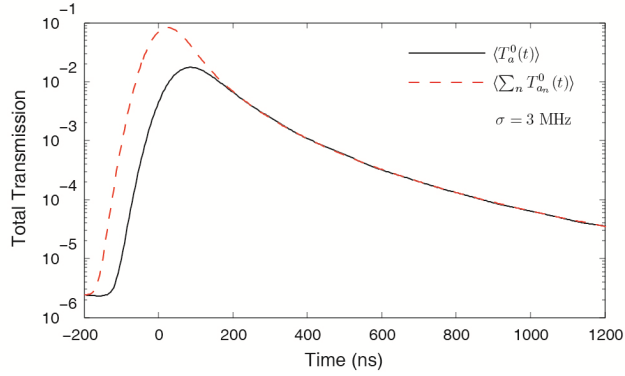

**Figure 4 | Dynamics of localized waves.** Semilogarithmic plot of ensemble average of pulsed transmission and incoherent sum of transmission due to all modes in random ensemble. The impact of absorption is eliminated as described in the text.

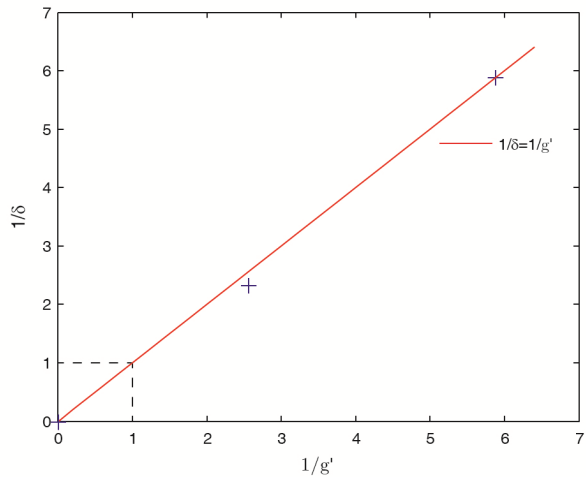

**Figure 5 | Localization parameters.** The degree of localization is tracked by the inverses of the mode overlap parameter, which is equivalent to the Thouless number, $\delta$, and the statistical conductance, $g'$, which reflects fluctuations in transmission. The dashed lines indicate the localization threshold at $\delta = g' = 1$. Measurements are made for sample lengths of 40 and 61 cm.

**Acknowledgements** This research was supported by the National Science Foundation (DMR0907285).

**Author Contributions** J. W. improved the apparatus, took the data, developed the modal decomposition and the time–frequency analysis algorithms, analyzed the data and contributed to writing the paper. A. Z. G. largely conceived and directed the research and wrote the paper.

**Author Information** Reprints and permissions information is available at ww.nature.com/reprints. The authors declare no competing financial interests. Readers are welcome to comment on the online version of this article at www.nature.com/nature. Correspondence and requests for materials should be addressed to J. W. (jing.wang@qc.cuny.edu) or A. Z. G. (genack@qc.edu).

## Methods

**Experimental method.** Spectra of the microwave field transmission coefficient through random waveguides are taken with use of a vector network analyzer. The sample is a copper tube filled with randomly positioned 99.95% alumina spheres of diameter 0.95 cm and refractive index 3.14 embedded in Styrofoam spheres of diameter 1.9 cm and refractive index 1.04 to achieve an alumina filling fraction is 0.068. The tube is 99.999% copper and highly reflective to minimize losses in reflection. The tube has a diameter of 7.2 cm and supports 30 transverse propagating modes over the frequency range of the experiment of 10-10.24 GHz. The sample is locally three dimensional within the reflecting tube with length much greater than its diameter. The random waveguide is thus similar to a microscopically disordered wire in electronics. The wave incident at different parts of the input surface of this quasi-one-dimensional sample is thoroughly mixed on the output surface and intensity statistics are essentially uniform across the output apart for fluctuations arising from the finite number of transverse modes. Radiation is strongly scattered within the sample over the frequency range of the experiment which is just above the first Mie resonance of the alumina spheres27. The value of $\delta$ varies by less than 10% over this frequency range in the sample with $L = 61$ cm.

The signal field output of the network analyzer is amplified before it is directed at the sample through a horn on axis with the sample tube placed as shown in Fig. 1. The transmitted field is detected by a wire antenna translated on a 2-mm square grid over the output surface. The antenna is 4 mm long and 300 μm in diameter and is connected to the network analyzer through a cable. The signal detected is the integral over the volume of the antenna of the component of the field directed along the wire axis. The dimensions of the wire are much smaller than the microwave wavelength in air of ~ 3 cm. Speckle patterns at each frequency are obtained from field spectra taken at each point on the output surface. The speckle patterns shown in Figs. 1 and 2 are obtained using the two-dimensional Whittaker-Shannon sampling theorem[31]. Ensembles of sample realizations are created by rotating the sample tube for several seconds after each measurement.

**Modes of an open system.** The field of a mode of an open system may be factorized into space and time varying terms after pulse excitation. The time varying term for the nth mode is $\cos\omega_n t \exp(-\frac{\Gamma_n}{2}t)$ for t>0 and zero for t<0. The decay rate of energy for the nth mode is $\Gamma_n$. The Fourier transform of the single sided exponential decay is the factor, $\varphi_n(\omega) = \frac{\Gamma_n/2}{\Gamma_n/2 + i(\omega - \omega_n)}$. Since the form of $\varphi_n(\omega)$ is independent of dimension, we expect the influence of absorption upon the mode linewidth and strength, similarly, will also not depend on dimension. We therefore explored the impact of absorption in a simpler 1D geometry using the transfer matrix simulations. We found the product of $a_n\Gamma_n$ is independent of absorption and that spectra found without absorption are replicated to high accuracy by substituting $a_{n,j}(\mathbf{r})\Gamma_n/\Gamma_n^0$ for $a_{n,j}(\mathbf{r})$ and $\Gamma_n^0 = \Gamma_n - 1/\tau_a$ for $\Gamma_n$ in equation (1). This substitution was used to find the spectra that would be obtained in the equivalent nonabsorbing samples. The absorption rate, $1/\tau a$, is determined from measurements of the energy decay rate in a sample with reflecting end caps in which leakage from the sample is suppressed. We expect the spatial modes of the medium are orthogonal over the volume of the sample28 but not across the output plane. This makes it possible for different modes to have similar speckle patterns.

**Separable nonlinear least square fitting.** The separable nonlinear least squares fitting method was used to simultaneously decompose field spectra at each of 45 positions into a sum over modes using a common set of $\omega_n$, $\Gamma_n$. The minimum is found for the sum, $\min \chi^2 = \min \sum_{p=1}^{P=45} \left| E(r_p, \nu) - \sum_{n=1}^{N} a_n(r_p)\varphi_n \right|^2$, where $r_p$ are the different positions whose spectra are fit. The problem is solved to yield the $\omega_n$, $\Gamma_n$ and the mode speckle patterns by using classical nonlinear least squares optimization methods, such as the Levenberg-Marquardt algorithm. In matrix format, this is,

$$\min \chi^2 = \min \|\mathbf{E} - \mathbf{\Phi a}\|^2, \qquad (2)$$

where each column of the matrix $\mathbf{E}$ represent a spectrum $\text{Re}E(v_1)$, $\text{Im}E(v_1)$, …, $\text{Re}E(v_M)$, $\text{Im}E(v_M)$ at a single position $r_p$ and each column of $\mathbf{a}$ represents the corresponding complex amplitudes of the modes, $\text{Re}a_1$, $\text{Im}a_1$, …, $\text{Re}a_N$, $\text{Im}a_N$ at position $r_p$. The elements of matrix $\mathbf{\Phi}$ are, $\Phi_{2m-1,2n-1} = \text{Re } \varphi_n(v_m)$, $\Phi_{2m-1,2n} = -\text{Im } \varphi_n(v_m)$, $\Phi_{2m,2n-1} = \text{Im } \varphi_n(v_m)$ and $\Phi_{2m,2n} = \text{Re } \varphi_n(v_m)$ for the $n^{th}$ mode at $m$ frequency point. The matrix $\mathbf{a}$ can very nearly be expressed as, $\mathbf{a} = \mathbf{\Phi}^\dagger \mathbf{E}$, where $\mathbf{\Phi}^\dagger$ is the Moore-Penrose pseudo-inverse of the matrix $\mathbf{\Phi}$[32,33]. Eliminating the linear coefficient $\mathbf{a}$, the fitting problem becomes,

$$\min \chi^2 = \min \|(\mathbf{I} - \mathbf{\Phi}\mathbf{\Phi}^\dagger)\mathbf{E}\|^2, \qquad (3)$$

where $\|...\|$ indicate the Euclidean norm of the matrix and $\mathbf{I}$ is the unit matrix. The problem can be solved using classical nonlinear least squares optimization methods, such as the Levenberg-Marquardt algorithm. Fitting was performed simultaneously for $P = 45$ points. The set of $\omega_n$ and $\Gamma_n$ obtained from this global fit were then used to solve for the spatial mode amplitudes $a_{n,j}(r_p)$ for all measured points.

**Time-frequency analysis.** In order to reveal the impact of modes on the dynamics of waves transmitted through random media, we compute the temporal evolution of the field as a function of the carrier frequency, $v_0$, of an incident Gaussian pulse, The spectrum of the incident pulse, $E_0(r, t, v_0) \sim \exp(-(t^2/2\sigma_t^2))\cos(2\pi v_0 t)$ is a Gaussian in the frequency domain, $E_0(t) \sim \exp(-(v-v_0)^2/2\sigma^2)$, where $\sigma = (2\pi\sigma_t)^{-1}$. Figures 3 and 4 are computed from the measured field spectra using time-frequency analysis[34] for pulses with $\sigma = 0.85$ MHz and 3 MHz, respectively, corresponding to Gaussian temporal pulse widths of $\sigma_t = 187.4$ ns and 53.1 ns. The increasing impact of long-lived narrow-line modes is seen in the time-frequency spectra in Fig. 3. The narrow linewidth of the incident pulse makes it possible to accurately determine the central frequencies of long-lived modes. These are used as initial values of parameters in the fit of the mode expansion in equation (1) to the measured spectrum.